\documentclass[aps,prb,twocolumn,showpacs,amsmath,amssymb]{revtex4}
\usepackage{epsfig}
\usepackage{graphicx}
\usepackage{booktabs}
\usepackage{multirow}
\usepackage{tabularx}
\usepackage{times}
\usepackage{color}
\usepackage{subfigure}

\definecolor{Red}{rgb}{1,0,0}

\definecolor{Blu}{rgb}{0,0,01}

\definecolor{Green}{rgb}{0,1,0}

\newcommand{\be}{\begin{equation}}
\newcommand{\ee}{\end{equation}}
\newcommand{\e}{\mathrm{e}}

\newcommand{\D}{\Delta}
\begin{document}

\title{Spin-flip scattering and critical currents in ballistic half-metallic $d$-wave Josephson junctions}

\author{Henrik Enoksen$^{1}$, Jacob Linder$^{1}$ and Asle Sudb{\o}$^{1}$}

\affiliation{$^{1}$ Department of Physics, Norwegian University of
Science and Technology, N-7491 Trondheim, Norway}
 
\begin{abstract}
We analyze the dc Josephson effect in a ballistic superconductor/half-metal/superconductor junction by means of the Bogoliubov--de Gennes equations. We study the role of spin-active interfaces and compare how different superconductor symmetries, including $d$-wave pairing, affect the Josephson current. We analyze the critical current as a function of junction width, temperature, and spin-flip strength and direction. In particular, we demonstrate that the temperature-dependence of the supercurrent in the $d_{xy}$-symmetry case differs qualitatively from the $s$- and $d_{x^2-y^2}$-symmetries. Moreover, we have derived a general analytical expression for the Andreev bound-state energies which shows how one can either induce $0-\pi$-transitions or continuously change the ground state phase of the junction by controlling the magnetic misalignment at the interfaces. 
\end{abstract}
 
\date{\today}
\pacs{72.25.-b,73.20.-r,73.20.At,73.40.-c,74.50.+r,75.70.Cn,85.25.Cp}

\maketitle

\section{Introduction}

The antagonistic nature of superconductivity and ferromagnetism makes their coexistence an unlikely one in bulk materials. In conventional superconductors, the current is carried by Cooper pairs consisting of two electrons in a spin singlet state, i.e. with anti-parallel spins.~\cite{BCS} Ferromagnets on the other hand, favor parallel spin-alignment. Hence, one might expect that the proximity-effect arising when bringing a ferromagnet into contact with a superconductor would decay rapidly due to this antagonistic nature.

However, it has been predicted that this rapid decay may not occur if there is some magnetic inhomogeneity present at the interface between a superconductor and a ferromagnet. Bergeret, Volkov and Efetov~\cite{bergeret} proposed a theory that coherent triplet Cooper pairs can be induced in ferromagnets, which in turn gives rise to a long-range triplet current. Several experiments have detected signs of this long range triplet current,~\cite{robinson,wang,khaire,sosnin,sprungmann} and it is evident that a magnetic inhomogeneity is causing triplet paring with parallel spins. {{Long range triplet currents should also appear in inhomogeneous magnetic  junctions with $d$-wave superconductors.~\cite{volkov_prl_09}}}

Half-metals~\cite{pickett} are fully spin-polarized ferromagnets, i.e. they are metallic for one spin-direction and insulating for the other. Some half-metals include CrO$_2$, La$_{0{.}7}$Sr$_{0{.}3}$MnO$_3$, and Fe$_3$O$_4$.~\cite{coey} Keizer \textit{et al.}~\cite{keizer} were the first to report long-range supercurrents through CrO$_2$. However, this group reported large variations in the magnitude of the critical current in the different samples. Moreover, some samples did not show any long-ranged supercurrent at all. Recent experiments have detected stable long range currents believed to be caused by a formation of spin-triplet pairing.~\cite{aarts} All experiments report some sort of spin-flip scattering caused by a magnetic inhomogeneity is required to obtain the long-range currents. 

As there is only one spin direction at the Fermi level in half-metals, the current passing through will be completely spin-polarized. This has the potential for useful applications in low-temperature nanoelectronics. There have been several theoretical works on half-metallic Josephson junctions,~\cite{eschrig,zheng,eschrig_nphys_08, asano_prl_07, galaktionov_prb_08, kupferschmidt_prb_11} but to our knowledge no-one has considered $d$-wave superconductors {{in this context}}. From a more general perspective, the interplay between unconventional superconductivity, such as $d$-wave pairing, and half-metallicity has been studied in the context of magnetoresistance and spin-injection properties of cuprate/manganite hybrid structures \cite{jakob_apl_95, pena_prl_05, nemes_prb_08, salafranca_prl_10}. The study of a half-metallic $d$-wave Josephson junction therefore has  relevance for these types of materials.

In this paper, we will study how the supercurrent through unconventional half-metallic Josephson junctions depends on the properties of spin-active interfaces. We also study how the critical current depends on parameters like junction width and temperature for different levels of spin flip scattering. Comparing the different superconducting symmetries, we find that a $d_{xy}$-wave superconductor gives a somewhat different result than $s$-wave and $d_{x^2-y^2}$-superconductors. Junctions made of of half-metals and the latter two types of superconducting order show a non-monotonic temperature dependence of the critical current, while corresponding junctions involving $d_{xy}$-superconductors do not show this behavior. Moreover, the junctions involving $d_{xy}$-superconductors may sustain a critical current which is considerably larger than the two other superconducting symmetries. We also show that one can induce $0-\pi$-transitions~\cite{ryazanov} by switching the alignment of the magnetic interface fields from parallel to anti-parallel. It is also possible to tune the ground state superconducting phase by rotating the magnetic misalignment field at one interface relative to the other interface. These results are obtained numerically. We also provide a general analytical expression for the Andreev bound-state (ABS) energy spectrum which confirms our numerical findings.

The rest of the paper is organized as follows. In section II, we introduce our model and notation. Our results are presented and discussed in section III, and we present our conclusions in section IV. We use $\hat{\dotso}$ for $4 \times 4$ matrices, and $\underline{\dotso}$ for $2 \times 2$ matrices. Boldface notation is used for 3-vectors.

\section{Model and formalism}
\begin{figure}
  \centering
  \includegraphics[width=\columnwidth]{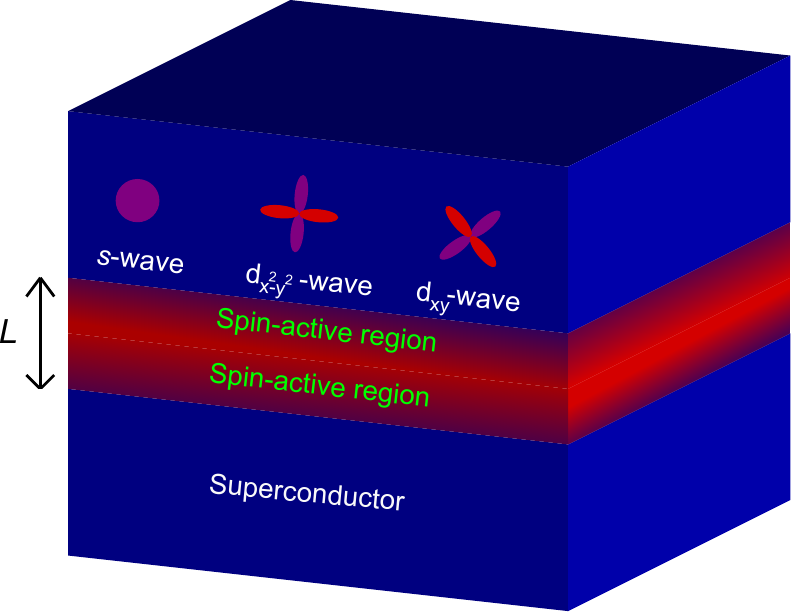}
  \caption{Schematic illustration of our system. A half-metal of width $L$ is sandwiched between two superconductors which are considered as reservoirs. The superconductors may have either an $s$-wave or $d$-wave symmetry. The interface regions are assumed to be spin-active due to e.g. magnetic disorder or misaligned local magnetic moments.}
  \label{fig:oppsett}
\end{figure}
We use a modified BTK (Blonder-Tinkham-Klapwijk) theory~\cite{btk} which takes into account arbitrary pairing symmetries and a spin-active S/F interface. We consider an S/F/S junction of width $L$, see Fig.~\ref{fig:oppsett}. The spin-active interfaces may induce long-range triplet 
currents. The propagation of quasiparticles is described by the Bogoliubov-de Gennes equations
\be
\hat{H}\psi = \varepsilon\psi
\ee
where $\psi$ is the eigenstate with energy eigenvalue $\varepsilon$. The Hamiltonian of the system is given by
\begin{equation}
\hat{H} = \begin{pmatrix}
\underline{H_0} + \underline{V} & i\D(\theta)\underline{\sigma_{2}} \\
-i\D(\theta)^{*}\underline{\sigma_{2}} & -\underline{H_0} - \underline{V}^{*}\\
\end{pmatrix} 
\end{equation}
where 
\begin{align}
 \underline  {H_0} =& \left(-\frac{\nabla^2}{2m} - E_F\right)\underline{\sigma_0}  - h_z\Theta(x)\Theta(L-x)\underline{\sigma_3} \nonumber \\
\underline{V} =& \begin{pmatrix}
V_{\uparrow} & V_x - iV_y \\
V_x + iV_y & V_{\downarrow}
\end{pmatrix}\left[\delta(x) + \delta(x-L)\right].
\end{align}
Here, $\Theta(x)$ and $\delta(x)$ denote the Heaviside step-function and the delta-function, respectively. The $\sigma$-matrices are the Pauli matrices, $m$ is the effective mass of the quasiparticles in both the superconductors and the ferromagnet and $E_F$ is the Fermi energy. We assume equal Fermi energies in the different regions of the junction. The superconducting gap is denoted by $\Delta(\theta) = \Delta_0(T)g(\theta)[e^{i\varphi_L}\Theta(-x) + e^{i\varphi_R}\Theta(x-L)]$, where $g(\theta)$ accounts for the superconducting pairing symmetry, $\Delta_0(T)$ is the temperature dependent gap amplitude and $\varphi_{L(R)}$ is the phase of the left(right) superconductor. We consider the usual BCS temperature dependence $\Delta_0(T) = \Delta_0\tanh\left(1.74\sqrt{T_c/T-1}\right)$ where $T_c$ is the superconducting critical temperature. The exchange energy is $h_z$, and its direction is parallel with the $z$-axis. We will consider the limit where the ferromagnet becomes half-metallic, i.e. $h_z \rightarrow E_F$. 

\begin{figure}
  \centering
  \includegraphics[scale=0.7]{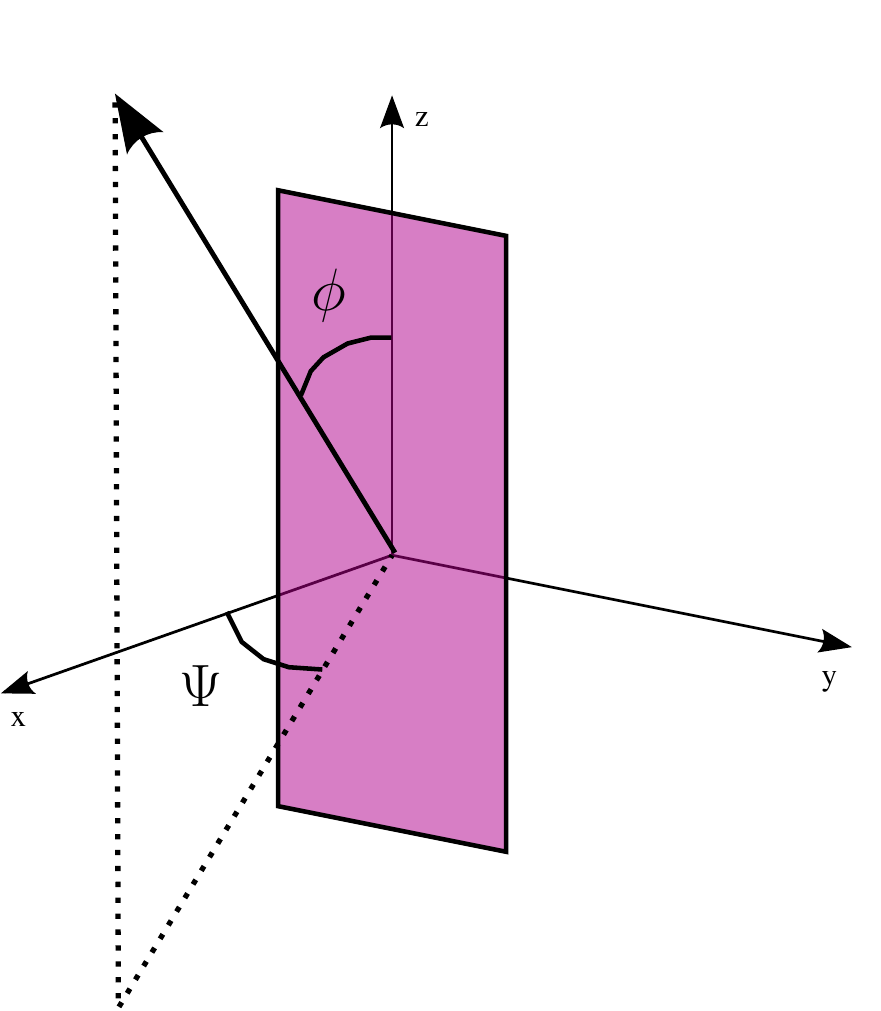}
  \caption{The barrier magnetic moment at the interface and its misalignment angles $\Psi$ and $\phi$. The bulk magnetization in the ferromagnet is assumed to be aligned with the $z$-axis.}
\label{fig:interface}
\end{figure}
The barrier magnetic moment constitutes a spin-dependent potential $\mathbf{V} = (V_x,V_y,V_z)$, where the components are
\begin{align}
V_x =& -\rho V_0 \cos\Psi\sin\phi\nonumber \\
V_y =& -\rho V_0 \sin\Psi\sin\phi \nonumber \\
V_z =& -\rho V_0 \cos\phi
\end{align}
and $V_{\sigma} = V_0 + \sigma V_z$. $\sigma = \pm 1$ for spin-up and spin-down. See Fig.~\ref{fig:interface} for an illustration of the barrier magnetic moment and its misalignment angles $\Psi$ and $\phi$. The non-magnetic barrier potential is $V_0$, while $\rho$ is the ratio between the magnetic and non-magnetic potentials, i.e.
\begin{equation}
  \rho = |\mathbf{V}|/V_0.
\label{eq:rho}
\end{equation}
We assume that the bulk magnetization of the ferromagnet is aligned with the $z$-axis in Fig.~\ref{fig:interface}. For $\rho \neq 0$ with $\phi = 0$ we have spin-mixing and with $\phi \neq 0$ we also have spin-flip. For details on spin-mixing and spin-flip, see e.g. Ref.~\cite{eschrig}

Solving the BdG-equations yields the wavefunction in the different regions of our system.\cite{jacobHM} We can have have four different incoming quasiparticles, electron like quasiparticles (ELQ) with spin up and down, and hole-like quasiparticles (HLQ) with spin up and down. For an incident spin-up electron in the left superconductor, the wave function is
\begin{align}
\psi_L(x) =& [u_\mathrm{L}(\theta_{+}),0,0,v_\mathrm{L}(\theta_{+})\e^{-i\gamma^{+}_{\mathrm{L}}}]\e^{ik_+x}  \nonumber \\
+& r_e^{\uparrow}[u_\mathrm{L}(\theta_{-}),0,0,v_\mathrm{L}(\theta_{-})\e^{-i\gamma^{-}_{\mathrm{L}}}]\e^{-ik_+x} \nonumber \\
 +& r_e^{\downarrow}[0, u_\mathrm{L}(\theta_{-}),\zeta v_\mathrm{L}(\theta_{-})\e^{-i\gamma^{-}_{\mathrm{L}}},0]\e^{-ik_+x} \nonumber \\
+& r_h^{\uparrow}[0, \zeta v_\mathrm{L}(\theta_{+})\e^{i\gamma^{+}_{\mathrm{L}}},u_\mathrm{L}(\theta_{+}),0]\e^{ik_-x} \nonumber \\
 +& r_h^{\downarrow}[v_\mathrm{L}(\theta_{+})\e^{i\gamma^{+}_{\mathrm{L}}},0,0,u_\mathrm{L}(\theta_{+})]\e^{ik_-x}.
\label{eq:psiL}
\end{align}
For this particular process, the coefficients $r_e^{\uparrow}, r_e^{\downarrow}, r_h^{\uparrow}, r_h^{\downarrow}$ describe normal reflection, normal reflection with spin-flip, novel Andreev reflection and usual Andreev reflection, respectively. 
We note that the momentum parallel to the interface is conserved for these processes.

The corresponding wave function in the right superconductor is
\begin{align}
\psi_R(x) =& t_e^{\uparrow}[u_\mathrm{R}(\theta_{+})\e^{i\varphi},0,0,v_\mathrm{R}(\theta_{+})\e^{-i\gamma^{+}_{\mathrm{R}}}]\e^{iq_+x} \nonumber \\
 +& t_e^{\downarrow}[0, u_\mathrm{R}(\theta_{+})\e^{i\varphi},\zeta v_\mathrm{R}(\theta_{+})\e^{-i\gamma^{+}_{\mathrm{R}}},0]\e^{iq_+x} \nonumber \\
+& t_h^{\uparrow}[0, \zeta v_\mathrm{R}(\theta_{-})\e^{i\gamma^{-}_{\mathrm{R}}}\e^{i\varphi},u_\mathrm{R}(\theta_{-}),0]\e^{-iq_-x} \nonumber \\
 +& t_h^{\downarrow}[v_\mathrm{R}(\theta_{-})\e^{i\gamma^{-}_{\mathrm{R}}}\e^{i\varphi},0,0,u_\mathrm{R}(\theta_{-})]\e^{-iq_-x},
\label{eq:psiR}
\end{align}
where the $t$'s are the transmission coefficients, corresponding to the reflection processes described above. We have defined $\e^{i\gamma_{\pm}} = \Delta(\theta_{\pm})/|\Delta(\theta_{\pm})|$ with $\theta_+ = \theta$ and $\theta_- = \pi - \theta$, and $\varphi = \varphi_R-\varphi_L$ is the phase difference over the junction. The parameter $\zeta$ accounts for singlet or triplet pairing in the superconductors. Here, we will consider only singlet pairing, i.e. $\zeta = -1$. Previous works have considered tunneling in $p$-wave superconductor/ferromagnet structures \cite{stefanakis_jcm_03, linder_prb_07, samokhin_prb_09, brydon_prb_11}. 
The coherence factors are defined as usual
\begin{align}
u(\theta) =& \sqrt{\frac{1}{2}\left( 1 + \frac{\sqrt{\varepsilon^2 - |\Delta(\theta)|^2}}{\varepsilon}\right)} \nonumber \\
v(\theta)  =& \sqrt{\frac{1}{2}\left( 1 - \frac{\sqrt{\varepsilon^2 - |\Delta(\theta)|^2}}{\varepsilon}\right)}.
\label{eq:coherencefactors}
\end{align}
$k_{\pm} = \sqrt{2m(E_F \pm \sqrt{\varepsilon^2 - |\Delta_L(\theta)|^2})}\cos\theta$ is the wavevector for ELQs ($k_+$) and HLQs ($k_-$) in the left superconductor and $q_{\pm}$ is the corresponding wavevectors in the right superconductor. The wave function in the half-metal is
\begin{align}
  \psi_{HM}(x) =& (e\e^{ik_e^{\uparrow}x} + f\e^{-ik_e^{\uparrow}(x-L)})[1,0,0,0] \nonumber\\
+& (e' \e^{ik_e^{\downarrow}x} + f'\e^{-ik_e^{\downarrow}(x-L)})[0,1,0,0] \nonumber \\
+& (g \e^{-ik_h^{\uparrow}x} + h\e^{ik_h^{\uparrow}(x-L)})[0,0,1,0]\nonumber \\
+& (g' \e^{-ik_h^{\downarrow}x} + h'\e^{ik_h^{\downarrow}(x-L)})[0,0,0,1],
\label{eq:psiHM}
\end{align}
where $k_{e,h}^{\sigma} = \sqrt{2m(E_F + \sigma h_z \pm \varepsilon)}\cos\theta$.

All scattering coefficients can be determined by matching wave functions at the interfaces. The boundary conditions are
\begin{align}
\psi_L(0^-) &= \psi_{HM}(0^+) \nonumber \\
\left.\partial_x[\psi_L(x) - \psi_{HM}(x)]\right|_{x=0} & = 2m\underline{V_L}\psi_{L}(0) ,
\label{eq:Lboundaryconditions}
\end{align}
for the left interface and
\begin{align}
\psi_{HM}(L^-) &= \psi_{R}(L^+) \nonumber \\
\left.\partial_x[\psi_{HM}(x) - \psi_{R}(x)]\right|_{x=L} & = 2m\underline{V_R}\psi_{R}(L) ,
\label{eq:Rboundaryconditions}
\end{align}
for the right. We will later use the parameter $Z = 2mV_0/k_F$ as a measure of interface transparency. Here, $k_F = \sqrt{2mE_F}$ is the Fermi wave vector. After the differentiation in the boundary conditions, we take the half-metallic limit $k_{e,h}^{\downarrow} \rightarrow 0$. From the boundary conditions we obtain a system of linear equations which yields the scattering coefficients. 

With the scattering coefficients at hand, we can use a generalized version of the Furusaki-Tsukuda~\cite{furusakitsukuda, tanakakashiwaya, asano, zheng} formalism to calculate the Josephson current 
\begin{align}
  I(\varphi) =& \frac{e}{4\beta}\sum_{\omega_n}\int\limits_{-\pi/2}^{\pi/2}\mathrm{d}\theta \Delta_L(\theta)\frac{k^+(\omega_n)+k^-(\omega_n)}{\Omega_n} \nonumber \\
& \times \left[\frac{a_1(\omega_n) - a_2(\omega_n)}{k^+(\omega_n)} + \frac{a_3(\omega_n) - a_4(\omega_n)}{k^-(\omega_n)}\right],
\label{eq:Ic}
\end{align}
where $\omega_n = (2n+1)\pi/\beta$ are fermionic Matsubara frequencies with $n = 0, \pm 1, \pm 2, ...$, and $\Omega_n = \sqrt{\omega_n^2+ |\Delta_L(\theta)|^2}$. $\beta$ is the inverse temperature. $k^+(\omega_n), k^-(\omega_n)$ and $a_i(\omega_n)$ are obtained from $k^+$, $k^-$ and $a_i$ by analytically continuing $\varepsilon$ to $i\omega_n$. $a_i$ with $i = 1,2,3,4$ are the ordinary Andreev reflection coefficients for incoming ELQ with spin-up or spin-down, and incoming HLQ with spin-up or spin-down, respectively. {{The summation over the Matsubara frequencies is performed numerically.}}

If we instead neglect the contribution from the incoming quasiparticle, we get a homogeneous system of linear equations
\begin{equation}
\Lambda \mathrm{x} = 0,
\label{eq:ABSsystem}
\end{equation}
where $\mathrm{x} = [r_e^{\uparrow},r_e^{\downarrow},r_h^{\uparrow},r_h^{\downarrow},t_e^{\uparrow},t_e^{\downarrow},t_h^{\uparrow},t_h^{\downarrow}]$ and $\Lambda$ is an $8 \times 8$ matrix obtained by expressing the scattering coefficients in the half-metal by the scattering coefficients in the left and right superconductor. By requiring a non-trivial solution of this system, $\det{\Lambda} = 0$, we find the Andreev bound state energy spectrum, $E_i$.

From the Andreev bound states we can find the Josephson current for a short junction $L/\xi \ll 1$, where $\xi$ is the superdonducting coherence length, in the ordinary way~\cite{golubovRevModPhys}
\begin{equation}
  I(\varphi) = 2e\sum_{i}{\frac{\partial E_i}{\partial \varphi} f(E_i)},
\label{eq:ABScurrent}
\end{equation}
where $e$ is the elementary charge and $f(E_i)$ denotes the Fermi-Dirac distribution function.
We find the critical current from $I_c = \max_{\varphi}|I(\varphi)|$.

\section{Results}
For conventional $s$-wave pairing we use $g(\theta) = 1$. For $d$-wave we use $g(\theta) = \cos(2\theta - 2\alpha)$ where $\alpha = 0$ and $\pi/4$ correspond to $d_{x^2-y^2}$- and $d_{xy}$-pairing, respectively. We will use the superconducting gap $\Delta_0$ as a unit of energy. The Fermi energy is $E_F = 1000\Delta_0$, the interface transparency is $Z = 1$ and we use $T = 0.2T_c$ unless otherwise stated. We consider equal gap amplitude for the different pairing symmetries to ease the comparison. Unless otherwise stated, we assume that the magnetic misalignment angles $\phi$ and $\Psi$ on both interfaces are equal.

\begin{figure}
\centering
\includegraphics[width=\columnwidth]{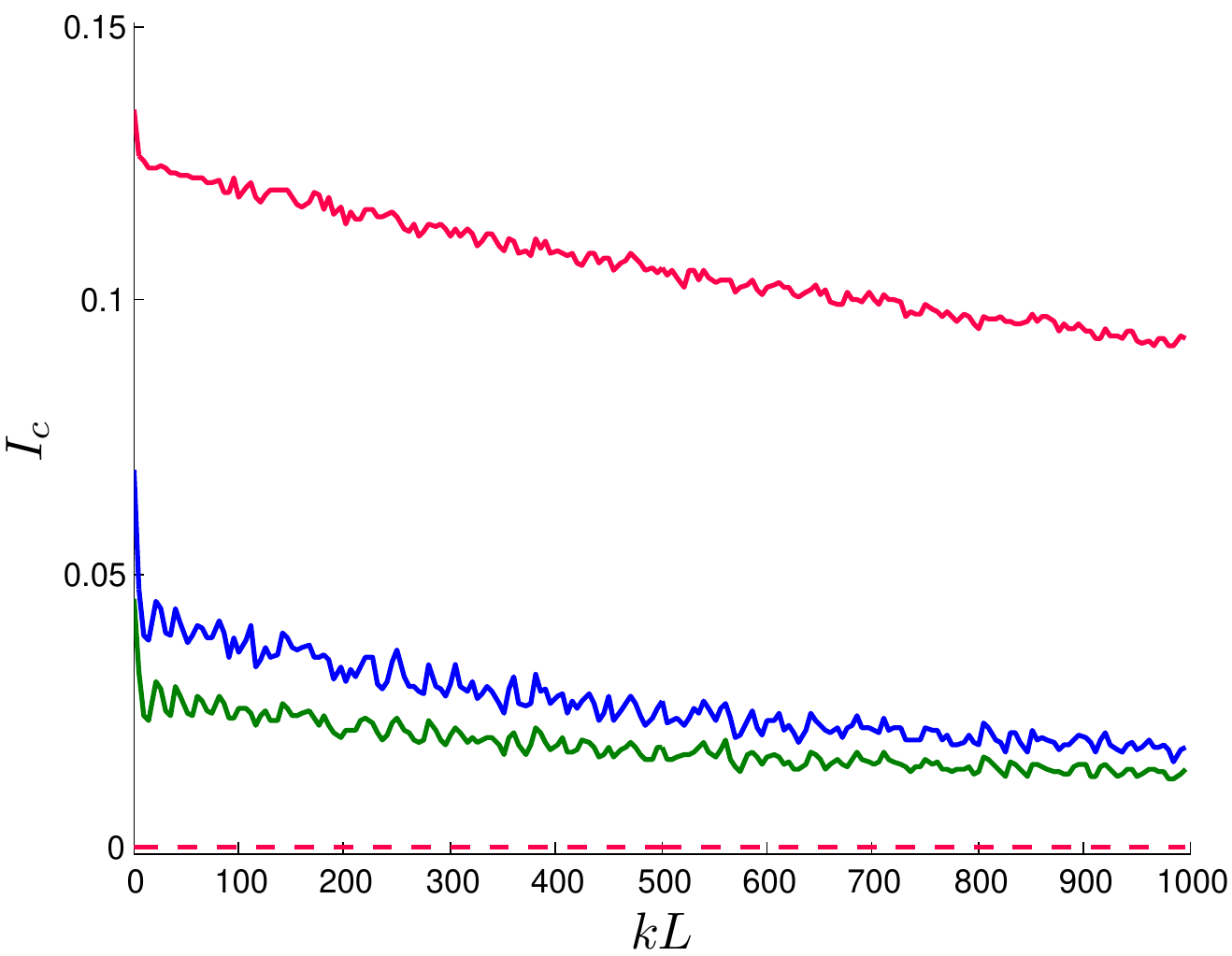}
\caption{Critical current as a function of junction width for $\rho = 0$ (dashed lines) and $\rho = 0.5$ (solid lines) with misalignment angles $\Psi = \pi/2$ and $\phi = \pi/2$. Blue, green, and red lines indicate $s$-, $d_{x^2-y^2}$-, and $d_{xy}$-wave pairing symmetry, respectively.}
\label{fig:criticalcurrentrho}
\end{figure}
Figure~\ref{fig:criticalcurrentrho} shows the critical current for all three pairing symmetries for two values of $\rho$ with misalignment angles $\phi = \pi/2$ and  $\Psi = \pi/2$. The first value, $\rho = 0$, corresponds to no spin-flip. As expected, there is no current in this case. However, when spin-flip scattering is present, $\rho = 0.5$, we see that a long-range current appears. We find that junctions with $s$- and  $d_{x^2-y^2}$-symmetry behave very similarly, with $d_{x^2-y^2}$ having a slightly lower current magnitude. The critical current is proportional to the gap amplitude multiplied by a weighting factor depending on the Andreev reflection coefficients (see Eq.~\eqref{eq:Ic}) averaged over the angles of incidence $\theta$. As the $s$-wave symmetry is independent of $\theta$, its average is larger than for the $d_{x^2-y^2}$. The $d_{xy}$-symmetry carries a current approximately three times larger than $s$- and $d_{x^2-y^2}$ for our parameter choice.

\begin{figure}
\centering
\includegraphics[width=\columnwidth]{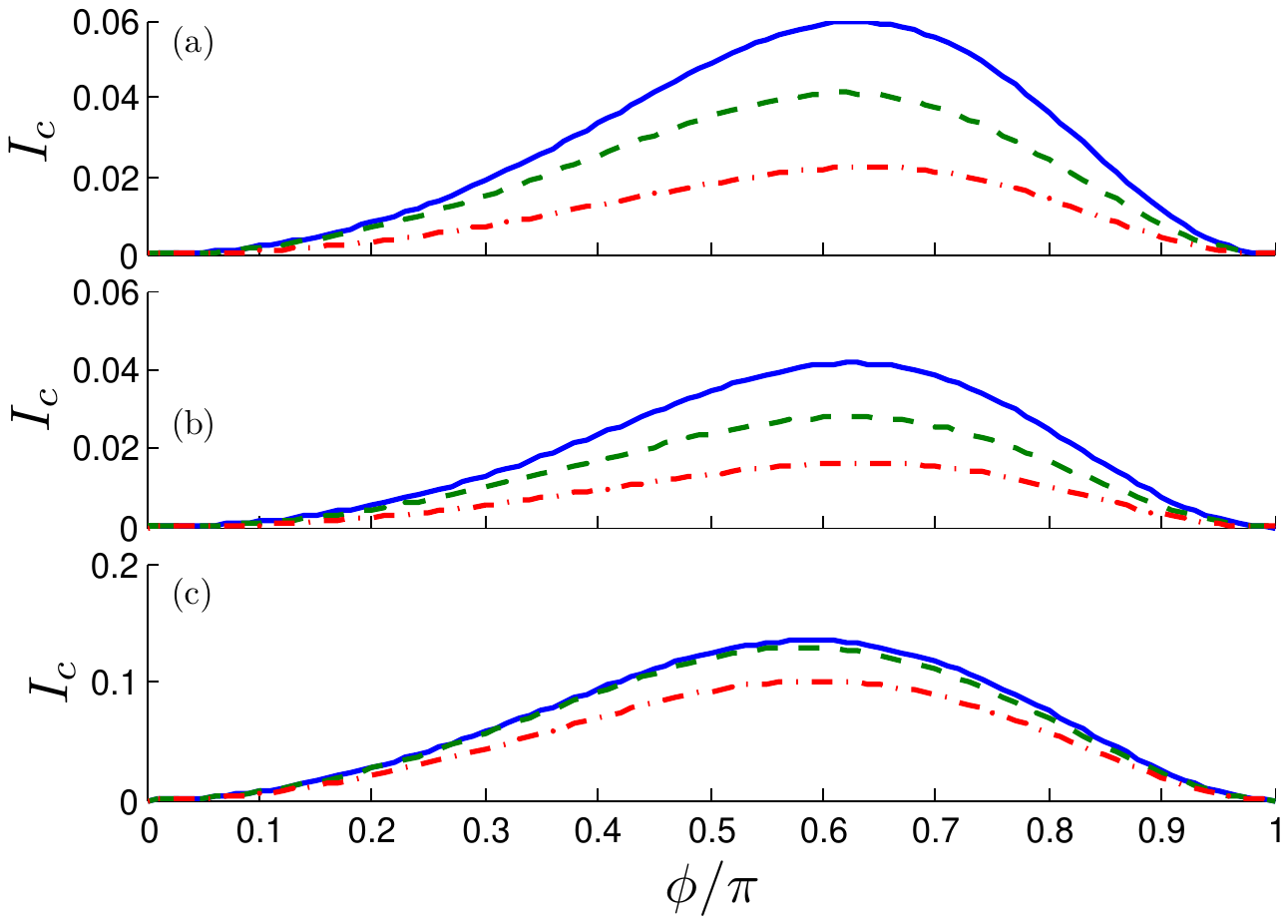}
\caption{Critical current as a function of the misalignment angle $\phi$. Solid, dashed and dash-dotted lines correspond to junction widths $kL=10, 100$, and $1000$, respectively. Panels (a)-(c) show $s$-, $d_{x^2-y^2}$-, and $d_{xy}$-wave pairing symmetry, respectively.}
\label{fig:Icphidep}
\end{figure}

Figure~\ref{fig:Icphidep} shows how the critical current depends on the magnetic misalignment angle $\phi$ with $\Psi = \pi/2$ for three junction widths $kL = 10, 100, 1000$ which corresponds to widths of approximately 2, 20 and 200 nm, respectively. Notice that we obtain the maximal critical current at a misalignment angle $\phi > \pi/2$ when the misalignment at both interfaces is parallel to each other. If we rotate the misalignment 180$^{\circ}$, which is equivalent to inverting the exchange field $h_z \rightarrow -h_z$, the maximum will appear at $\phi < \pi/2$. As long as $h_z \neq 0$, the system is not invariant under a spatial inversion.~\cite{linderalpha} Therefore, the direction of the exchange field relative to the misalignment field will determine whether the maximal current appears at misalignment angles smaller or larger than $\pi/2$.

\begin{figure}
\centering
\includegraphics[width=\columnwidth]{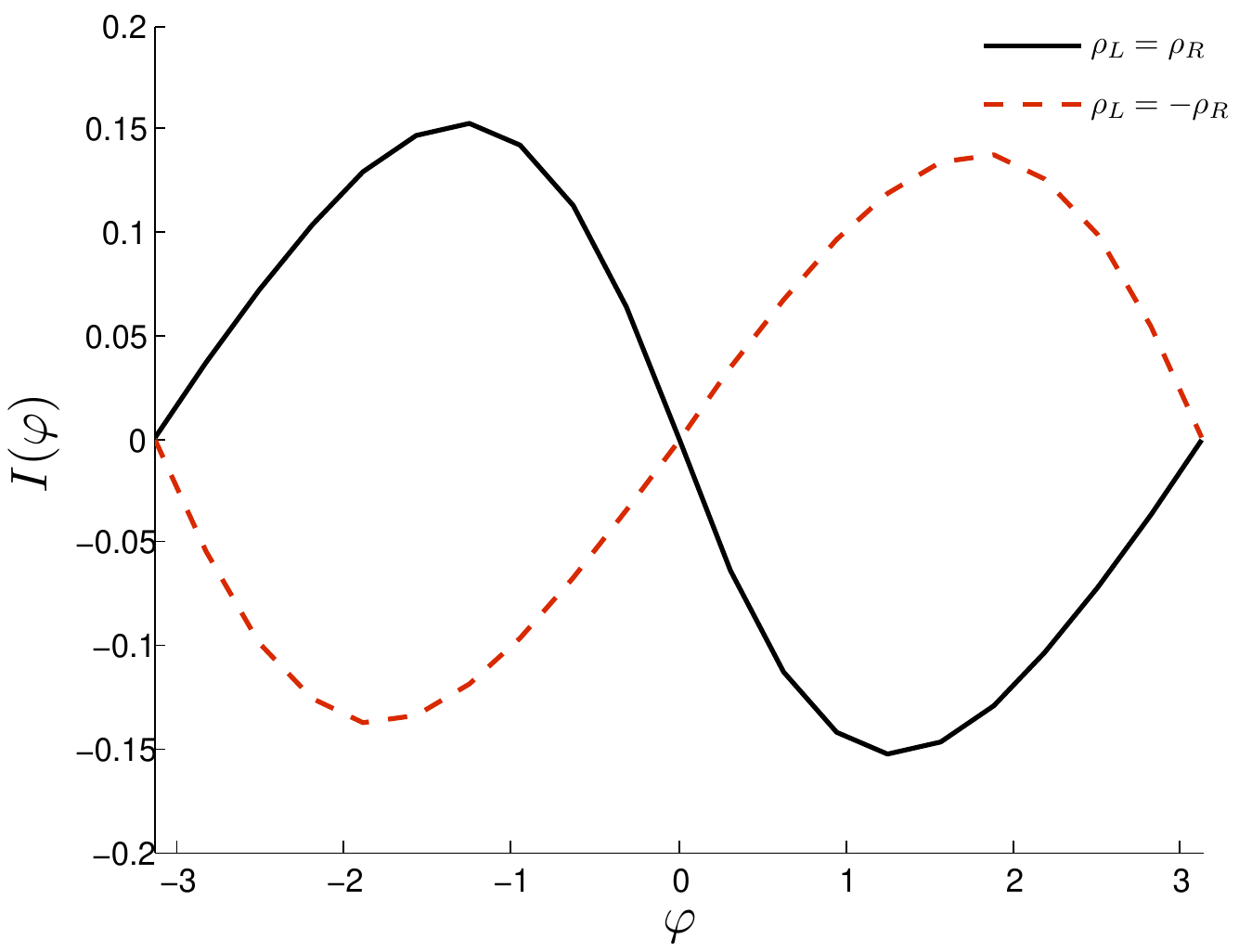}
\caption{Current-phase relation for a junction with $d_{xy}$-superconductors on both sides. The black solid line shows parallel misalignment at both interfaces, while the red dashed line shows anti-parallel misalignment.}
\label{fig:CPRrho}
\end{figure}

If the magnetic moments at the interfaces are switched from parallel ($\rho_L = \rho_R$) to anti-parallel alignment ($\rho_L = -\rho_R$), the current-phase relation changes sign as seen in Fig.~\ref{fig:CPRrho}. This indicates a $0-\pi$-transition. Hence, it is possible to induce $0-\pi$-transitions in half-metallic junctions by switching the relative direction between the two misalignment fields. Note that the other superconducting symmetries show the same behavior. We also note that for anti-parallel alignment, the maximal critical current is obtained at a misalignment angle $\phi = \pi/2$ for both interfaces.

\begin{figure}
\centering
\includegraphics[width=\columnwidth]{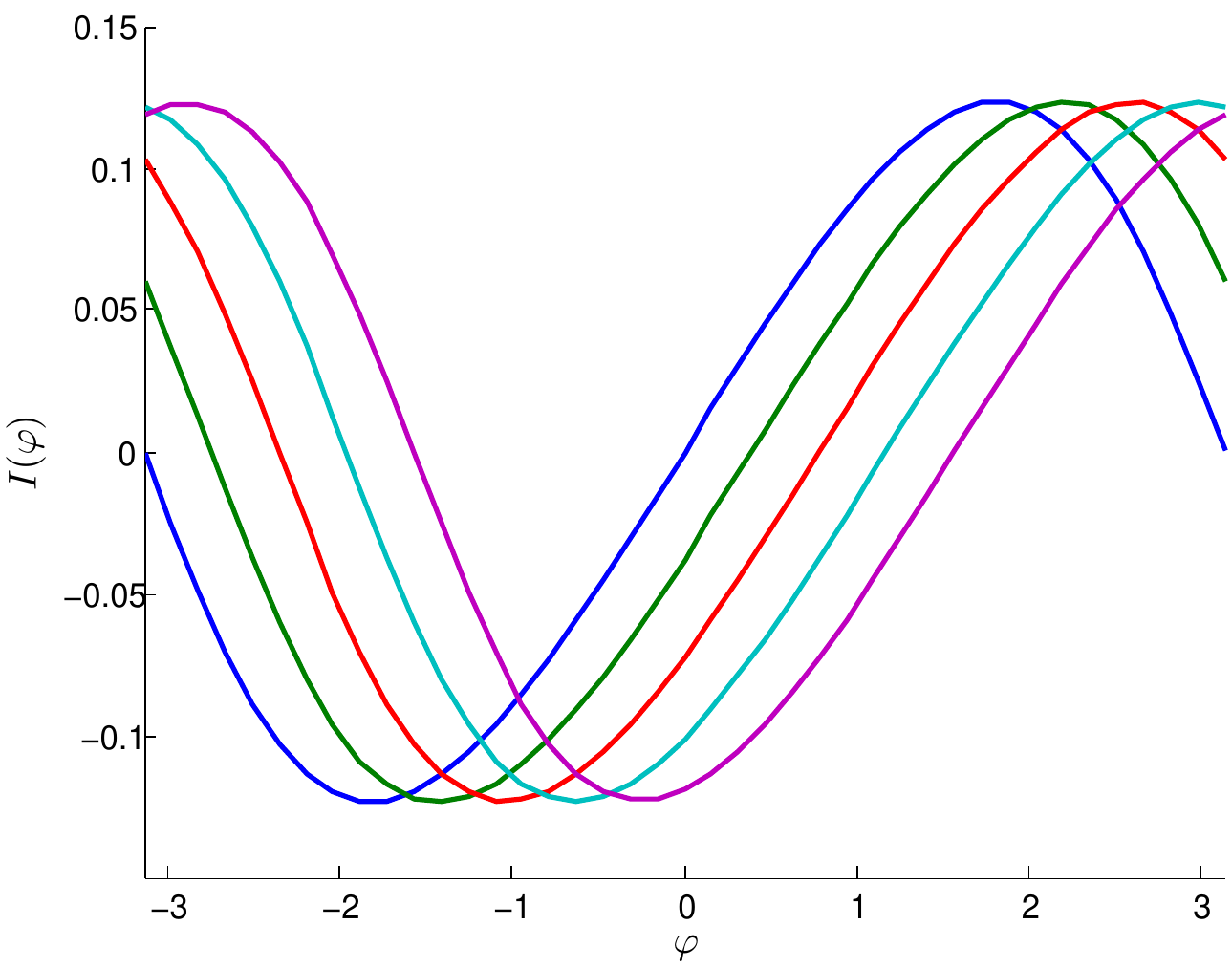}
\caption{Current-phase relation for a junction with $d_{xy}$-superconductors on both sides. The misalignment field in the $xy$-plane is rotated from parallel to 
perpendicular, i.e. $\Psi_{\mathrm{R}}$ changes from 0 to $\pi/2$ in steps of $\pi/8$ (left to right). $\Psi_{\mathrm{L}} = 0$ and $\rho_{\mathrm{L}} 
= \rho_{\mathrm{R}} = 0.5$. Here, $\phi_L = \phi_R = \pi/2$.}
\label{fig:cprsintocos}
\end{figure}

It is also possible to continuously change the ground state phase of the junction by rotating the misalignment field at one interface relative to the other. Fig.~\ref{fig:cprsintocos} shows the current-phase relation for a $d_{xy}$-junction when the magnetic moment at the left interface points in the $x$-direction while at the right interface the magnetic moment is rotated in the $xy$-plane from being parallel to the left interface, to pointing in the $y$-direction. Here, $\phi_L = \phi_R = \pi/2$. We see that the current changes continuously with $\Psi_R$. Hence, we can obtain a finite current at zero superconducting phase difference by tuning the magnetic misalignment fields. This is the same result as reported in Ref.~\cite{braude} for $s$-wave superconductors. We get the same results for $s$- and $d_{x^2-y^2}$-wave junctions.

To gain further insight into the physical mechanisms underlying our numerical results, we have solved Eq.~\eqref{eq:ABSsystem} analytically. This yields a rather large and complicated expression for the ABS-energies, but it is nevertheless possible to infer how the interplay between the misalignment angles and superconducting phase difference is manifested. The resulting expression is
\begin{widetext}
\begin{equation}
E = \pm\left|\Delta(\theta)\right|\sqrt{\frac{1}{2} + \frac{A_1 + A_2\tau\cos{(\varphi + \chi)} \pm \sqrt{B_1 + B_2\tau\cos{(\varphi + \chi)} + B_3\tau^2\cos^2{(\varphi + \chi)} }}{C}}\;,
\label{eq:generalABSenergy}
\end{equation}
\end{widetext}
where $A_i$, $B_i$ and $C$ are large expressions that depend on junction parameters like junction width $kL$, the barrier magnetic moment $\mathbf{V}$, the angle of incidence $\theta$ and the wave vectors $k_{e,h}^{\uparrow}$. $\tau = \alpha\rho^2V_0^2\sin^2{\phi_L}$ where $\alpha = \pm 1$ for parallel and anti-parallel misalignment, respectively, and $\chi = \Psi_R - \Psi_L$ is the difference between the azimuthal angle of the right and left interface magnetic moments. As the Jospehson current depends on the ABS-energy differentiated with respect to $\varphi$, see Eq.~\eqref{eq:ABScurrent}, much of the qualitative behavior can be explained with this expression. First, if we have no spin-flip, i.e $\phi = 0$ and/or $\rho = 0$, then $\tau = 0$. Every term containing $\varphi$ is multiplied by a $\tau$-factor, which shows that no spin-flip equals no current. This corresponds to the result we see in Fig.~\ref{fig:criticalcurrentrho} where the current vanishes without spin-flip.  Second, we see that a change of the magnetic misalignment on the two interfaces from parallel, $\alpha = 1$, to anti-parallel $\alpha = -1$ is the same as shifting $\varphi \rightarrow \varphi + \pi$, i.e. a $0-\pi$-transition like the one we see in Fig.~\ref{fig:CPRrho}. Lastly, the $\chi$-phase shows that it is possible to continuously change the ground state phase of the junction by tuning the interface magnetic moments as shown in Fig.~\ref{fig:cprsintocos}, since $\chi$ effectively renormalizes the superconducting phase difference: $\varphi\to\varphi+\chi$.

\begin{figure}
\centering
\subfigure[$s$-wave, $kL=100$]{
 \includegraphics[width=\columnwidth]{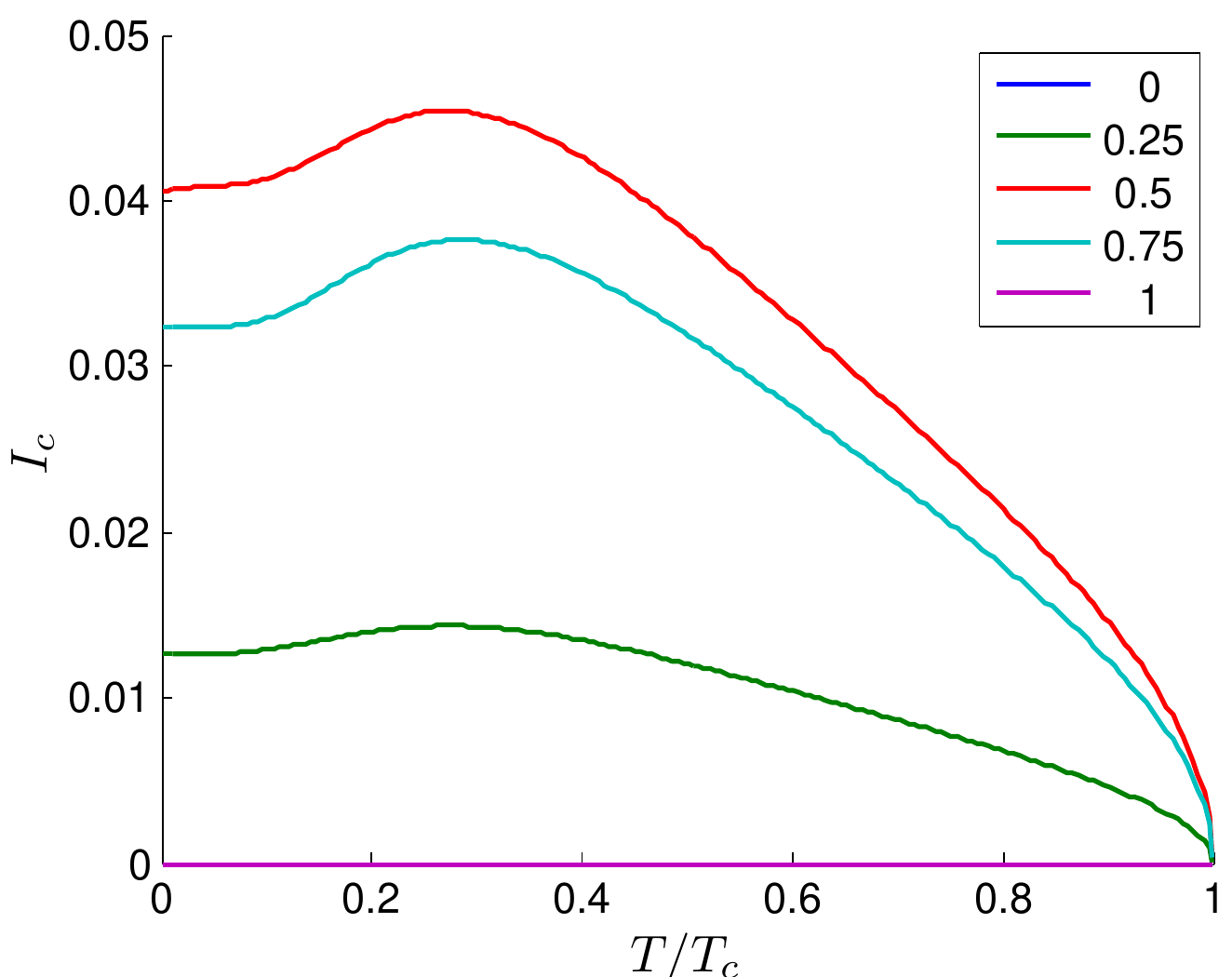}
 \label{fig:sIcT}
 }
\subfigure[$d_{x^2-y^2}$-wave, $kL=100$]{
 \includegraphics[width=\columnwidth]{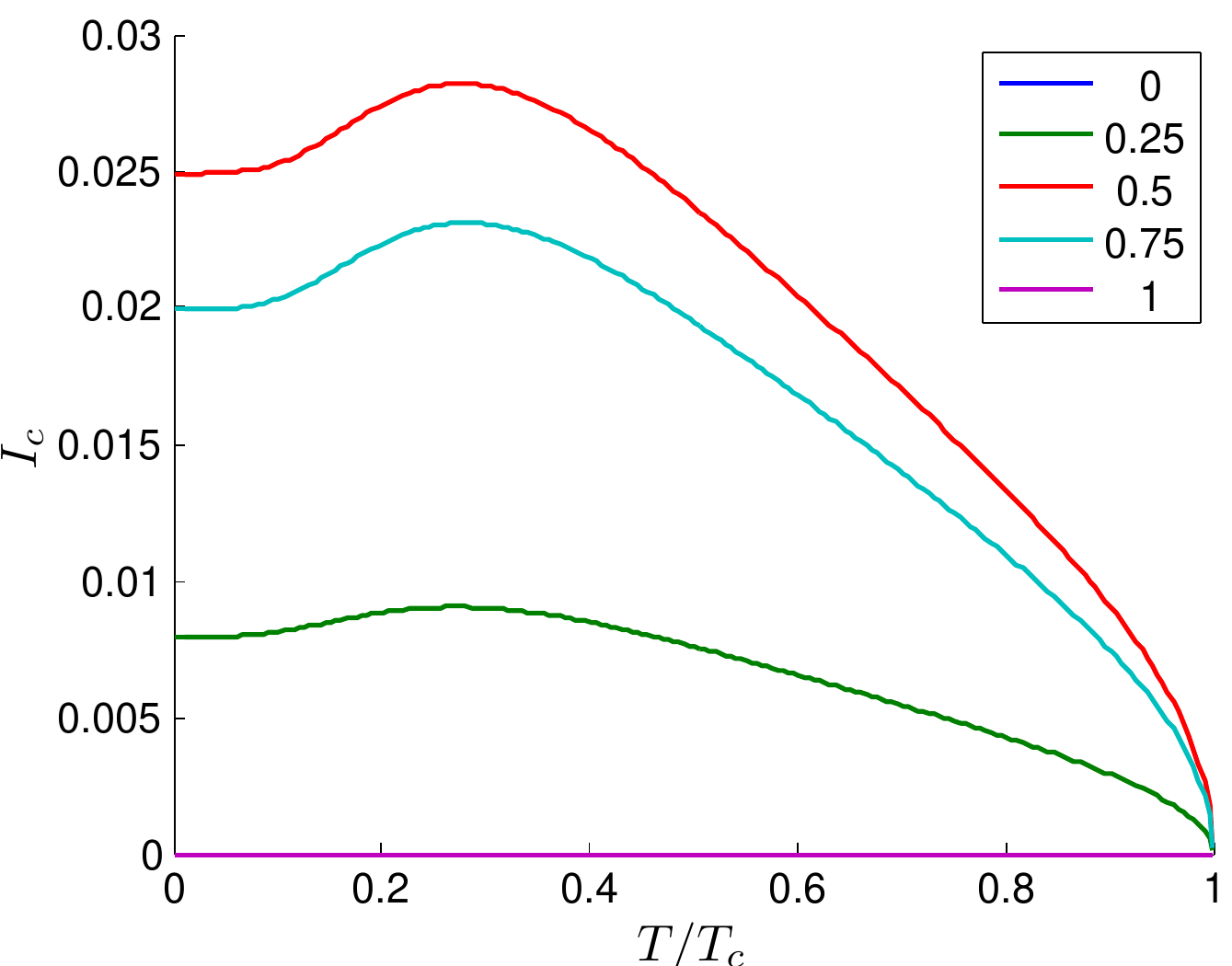}
 \label{fig:dx2y2IcT}
 }
\subfigure[$d_{xy}$-wave, $kL=100$]{
 \includegraphics[width=\columnwidth]{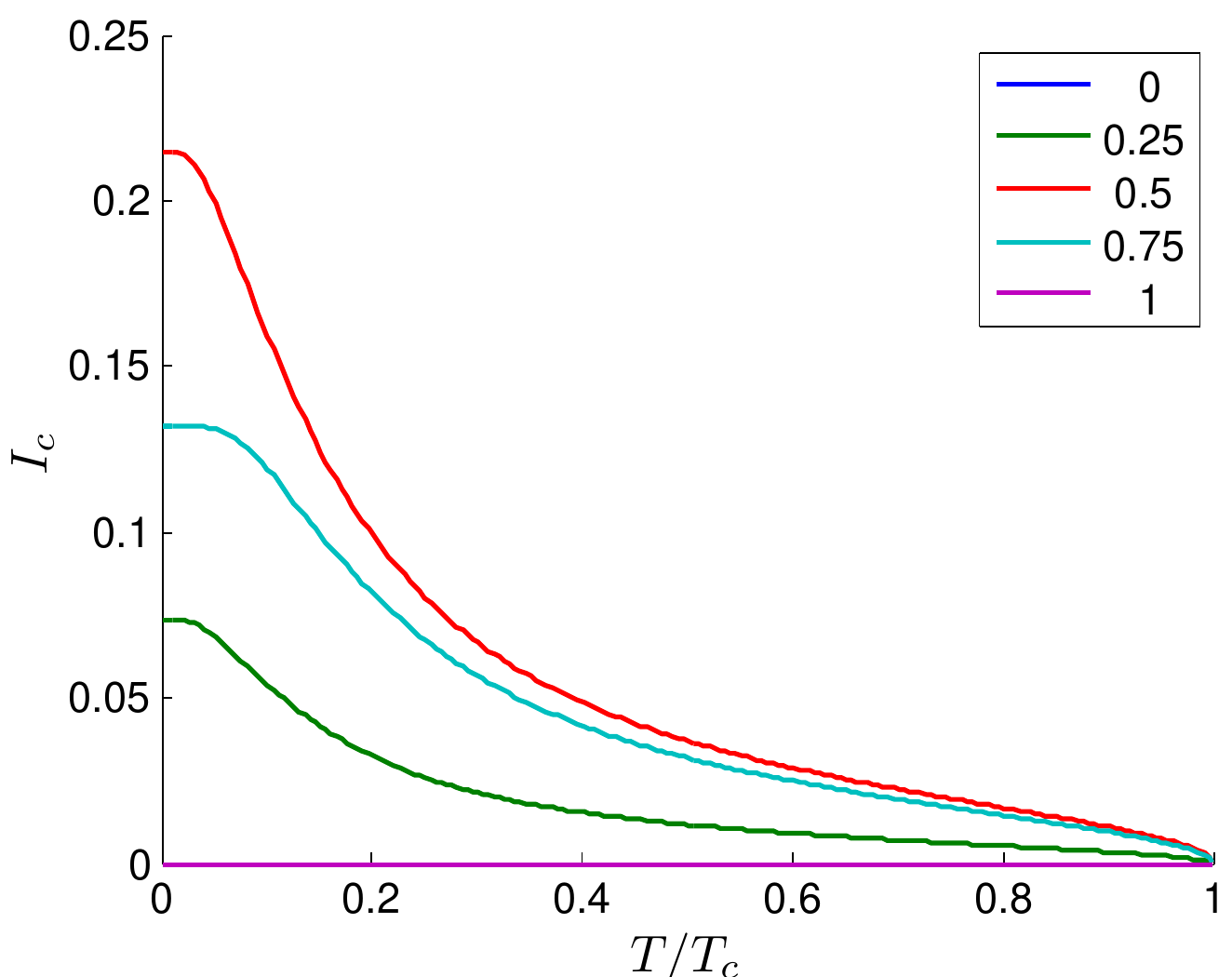}
 \label{fig:dxyIcT}
 }
\caption{Critical current as a function of temperature for all three superconducting symmetries at junction width $kL = 100$. Notice that the $d_{xy}$-symmetry decreases monotonically, while $s$- and $d_{x^2-y^2}$-symmetries have a non-monotonic behavior. The legends show the misalignment angle $\phi$ in units of $\pi$.}
\label{fig:IcT}
\end{figure}
The most significant difference between the three gap symmetries is the temperature dependence of the critical current. The Josephson current in the case of $d_{xy}$-symmetry decreases monotonically with increasing temperature, while the Josephson currents for the two other symmetries show a non-monotonic behavior. Figs.~\ref{fig:IcT} demonstrate the temperature dependence of the critical current for the different gap symmetries. 

This unusual temperature dependence can be explained by considering the proximity-induced density of states in the half-metal. The critical current is proportional to the energy integral of the density of states multiplied by the slope of the Fermi distribution:~\cite{DOS}
\begin{equation}
  I_c \sim \int{\mathrm{d}E\: N(E)\partial_{E}f(E)} \;,
\label{eq:Ic_integral}
\end{equation}
where $N(E)$ is a proximity-induced density of states in the half-metal, which vanishes when superconductivity vanishes.

As shown in Refs.~\cite{DOS,DOS2} the proximity-induced density of states for a half-metallic Josephson junction with $s$-wave superconductors has a peak at finite energies. Spin-flip processes at the interfaces create long-ranged spin-triplet pairs that penetrate the interface and create a peak in the density of states in the half-metal due to the formation of bound-states (in the ballistic limit).  This peak is the origin of the non-monotonic temperature behavior of the critical current which we also obtain, 
see Fig.~\ref{fig:sIcT}. Subgap bound-states may be thermally activated and contribute to the transport at a temperature $T^*$ of the 
order of the energies of the bound-states. Thus, when the temperature $T$ is increased towards $T^*$ the critical current can increase, although one should also take 
into account the fact that the superconducting order parameter is suppressed with increasing temperature. For the case of $s$-wave and $d_{x^2-y^2}$-wave superconducting 
gaps, the required temperature $T^*$ is of the order of the gap-amplitude. For 
a $d_{xy}$-symmetric gap, the bound-states tend to be at lower energies close to the Fermi surface \cite{jacobHM}, yielding a lower 
$T^*$, and hence a critical current which essentially is monotonically decreasing with temperature. We emphasize that all of this is 
contingent on the presence of spin-active tunneling barriers. Fig.~\ref{fig:dx2y2IcT} shows that $d_{x^2-y^2}$-pairing yields the 
same non-monotonic behavior as $s$-wave pairing, while Fig.~\ref{fig:dxyIcT} shows that $d_{xy}$-pairing decays in the usual 
monotonic way. 

{{We expect the same behavior for other superconducting symmetries as well: If the bound-states are at energies close to the Fermi level, i.e. the density of states peaks at or very close to $E_F$, we expect the usual monotonic temperature dependence. However, if the bound-states are at higher energies, i.e. the density of states peaks at higher subgap energies, we expect a non-monotonic temperature dependence.}}

\section{Conclusions}
In conclusion, we have analyzed the dc Josephson effect in an S/HM/S junction with spin-active interfaces. The possibility of spin-flip scattering at the interface creates equal spin triplet Cooper pairs in the half-metal, such that a long-range supercurrent may be sustained. We have  considered the role of unconventional superconducting symmetries, in particular $d$-wave, and found that $d_{xy}$-symmetry yields the largest critical current magnitude, while $s$- and $d_{x^2-y^2}$-symmetry show similar behavior. In addition, the temperature dependence of the critical
current is qualitatively different in the $d_{xy}$-wave case. Namely, while the current shows a non-monotonic behavior for $s$- and $d_{x^2-y^2}$-symmetry, it decays monotonically in the $d_{xy}$-wave case. This may be explained by considering the proximity-induced density of states in the half-metal. When the density of states in the half-metal has more weight near the Fermi level, as in the $d_{xy}$-wave case due to the hybridization of the bound-states at the S/HM interfaces, the temperature dependence of the critical current is monotonically decaying as the temperature increases. {{On the other hand, if the density of states has a peak far from the Fermi level, we expect a non-monotonic temperature dependence.}} We have also discussed how it is possible to switch between a $0$- and $\pi$-junction, both continuously and abruptly, by controlling the magnetic properties of the interfaces. Finally, we have found a general analytical expression for the Andreev bound-state energy spectrum which confirms our numerical results.

\acknowledgments

J. L. thanks M. Cuoco for useful discussions. J. L. and A. S. were supported by the Research Council of Norway, Grant No. 205591/V30 (FRINAT). 
H. E. acknowledges support from the Norwegian University of Science and Technology.

\end{document}